# Precise Parameters from Bayesian SED Fitting Indicate Thermally-Driven Mass Loss Likely Driver of Radius Valley

David Jordan[1,2] , Inseok Song[2]

## Abstract

Several planet formation models have been proposed to explain the gap in the population of planets between 1.8 $R_\oplus$ to 2.0 $R_\oplus$ known as the Radius Valley. To apply these models to confirmed exoplanets, accurate and precise host star and planet parameters are required to ensure the observed measurements correctly match model predictions. Previous studies have emphasized the need for a larger, more precise sample to further confirm dominant formation processes. By enhancing standard SED (Spectral Energy Distribution) fitting using Bayesian methods we derived highly accurate and precise host star and planet parameters. Specifically, we achieved median fractional uncertainties for stellar and planet radii of 2.4% and 3.4%, respectively. We then produced the largest, most precise sample to date of 1923 planets when compared to previous studies. This full sample, as well as a sampled filtered for host stellar masses between 0.8 and 1.2 $M_\odot$, are then used to derive the slope and position of the radius valley as a function of orbital period, insolation flux and stellar mass to compare them to predictive models and previous observational results. Our results are consistent with thermally-driven mass loss with a planet radius vs. orbital period slope of $R_p = P^{-0.142^{+0.006}_{-0.006}} e^{0.896^{+0.012}_{-0.010}}$ for the full sample leaning toward core-powered mass loss. The planet radius vs. insolation flux slope of $R_p = S_p^{0.136^{+0.014}_{-0.014}} e^{-0.085^{+0.031}_{-0.030}}$ for the filtered sample leaned toward photoevaporation. Also, the slope as a function of stellar mass for both samples appear more consistent with thermally driven processes when compared to models and previous studies.

---

[1] Corresponding author David.Jordan@uga.edu
[2] Department of Physics and Astronomy, The University of Georgia, Athens, GA 30602, USA

# 1. Introduction

Since the beginning of the Kepler mission, research efforts have greatly advanced our understanding of exoplanet formation and its dependency on various factors related to the planet host star. Using Radial Velocity monitoring data for 1266 stars, Asher Johnson et al. (2010) found that planet occurrence increases with stellar mass, characterized by a rise from 3% around M dwarfs (~0.5 $M_\odot$) to 14% around A stars (~2 $M_\odot$), at Solar metallicity. This points to a direct correlation between stellar mass and the fraction of stars with detectable planets and provided valuable insights into the planet-formation processes that guided future exoplanet search programs. As the Kepler mission started confirming more planets with the associated data on their host stars, many planets with radii between those of Earth and Neptune were discovered. Notably, these classes of planets do not exist in our solar system. Nonetheless, these intermediate planets are the most abundant in the Kepler sample of FGK type stars. This deviation raised an important question as to whether these sub-Neptune-size planets are primarily rocky or have low-density envelopes that significantly affect their size (Batalha et al. 2013).

Owen & Wu (2013) predicted a lack of large low-density planets close to the stars and a possible bimodal distribution in planet sizes with a deficit of planets around 2 $R_\oplus$. Lopez & Fortney (2013) emphasized the importance of core mass in planetary evolution and predicted that planets with radii between 1.8 and 4.0 $R_\oplus$ should become less common on orbits within 10 days. The impact of evaporation can be planetary mass dependent where giant planets do not lose H/He envelope while low-mass planet lose all their H/He envelope (Jin et al. 2014). Envelope mass-loss timescales vary non-monotonically with H/He envelope mass (Chen & Rogers 2016) and implies that ultra low-density planets can probably survive for Gyr timescales. These model predictions suggested an absence of planets between super-Earths greater than 1.5 $R_\oplus$ and sub-Neptunes less than 3.0 $R_\oplus$.

More accurate and precise host star radius measurements were necessary to confirm/reject the predicted absence of planets in the radius range of 1.5-3.0 $R_\oplus$. Once a precise stellar radius is known and the transit depth $\delta$ is measured, planet radius can be calculated, and uncertainties can be propagated. To improve the accuracy of both stellar and planetary radius measurements, careful consideration of observational techniques, data analysis methods, and incorporating additional sources of information to overcome observational challenges such as flux contamination, crowded field effect, and stellar activity is required (Mullally et al. 2015).

Techniques focused on overcoming the observational challenges included the California-Kepler Survey (CKS) (Petigura et al., 2017) which determined values for effective temperature ($T_{\text{eff}}$), and then used the Stephan-Boltzmann law, along with known luminosity measurements, to determine more precise stellar radii. This provided a substantial improvement in the median uncertainties of stellar radius, reducing them from 25% (Huber et al. 2014) to 11%. Subsequently, the median uncertainties in planet radii were reduced to 12%, exposing more detailed structures in the planet radius distribution. These more precise measurements further confirmed a gap in the radius distribution of small planets near their host stars. Subsequently, Fulton & Petigura (2018) used Gaia DR2 parallax data and the uncertainties of stellar radii measurements were reduced from 11% to 3%. Consequently, the uncertainties of planet radius measurements were thus reduced to 5% by incorporating Gaia astrometry data.

Core-powered mass loss and photoevaporation can produce a bimodal population of planet sizes consisting of either rocky cores or those with H/He envelopes. However, the primary difference lies in their expected dependence on stellar mass. It has been shown that core-powered mass loss is independent of stellar mass (Ginzburg et al. 2016), while photoevaporation predicts that sub-Neptunes should shift to lower insolation fluxes with decreasing stellar mass (Chen & Rogers 2016; Jin et al. 2014; Lopez & Fortney 2013; Owen & Wu 2013). Other proposed super-Earth formation mechanisms, including delayed formation in a gas-poor disk (Lee & Chiang 2016) and giant impacts stripping the envelope (Inamdar & Schlichting 2016), are found to be inconsistent with the initial observed results. It should be noted that these mechanisms are not mutually exclusive and could be operating simultaneously or at different times during planetary system formation so further investigation of this gap in planet populations around 2 $R_\oplus$ continued over the past several years.

Since the slope of the radius valley can provide insights into the formation and evolution of planetary systems (Fulton & Petigura 2018), analysis and measurement of the slope of the Radius Valley became a key aspect of several studies related to determining how the planet formation process shaped the gap. Lopez & Rice (2018) predicted that the transition radius between rocky and non-rocky planets should scale roughly as $R_{\text{evp}} \propto F_p^{0.11}$ (Insolation Flux), and $R_{\text{evp}} \propto P^{-0.15}$ (Orbital Period) for photoevaporation. This means the maximum size of bare-rocky planets should increase with increasing incident flux and with decreasing orbital period. For gas-poor formation, this model predicted that the transition radius between rocky and non-rocky planets should scale roughly as $R_p \propto F_p^{-0.08}$ (Insolation Flux), and $R_p \propto P^{0.11}$ (Orbital Period), indicating the maximum size of bare-rocky planets should decrease with increasing incident flux and with increasing orbital period if the lack of gas in the disk is driving the formation process. Cloutier & Menou (2020) suggested that thermally driven atmospheric mass loss may not be the dominant factor in the evolution of planets in the low stellar mass regime from their study of 328 planets around low mass M-dwarfs. Using 117 stars with accurate radii based on asteroseismic data (Lundkvist et al. 2016; Silva Aguirre et al. 2015), Van Eylen et al. (2018) claimed that the result is more consistent to the photoevaporation model than the late formation of gas-poor rocky planets. Gupta & Schlichting (2020) investigated the impact of the core-powered mass-loss mechanism (Ginzburg et al. 2016) on the radius valley and showed that the gradient of the valley is primarily governed by the atmospheric mass-loss timescales at the Bondi radius and shift in the radius valley to larger planet sizes around more massive stars. Martinez et al. (2019) confirmed the bimodal distribution of planetary radii with a gap at 1.9 $R_\oplus$ from the spectroscopic analysis of archival Keck/HIRES spectra. Berger et al. (2020) used 3308 planets under 4 $R_\oplus$ with Gaia DR2 data and arrived at a similar conclusion by Gupta & Schlichting (2020). Using more precise planetary radii for 970 planets, Petigura et al. (2022) found that the radius valley follows a power law relationship, consistent with predictions of X-ray, ultraviolet, and core-powered mass-loss mechanisms. They also showed no stellar mass dependency of the radius valley slope.

While most of the previous research as described above still pointed to thermally-driven mass loss, it was evident that both XUV photoevaporation and core-powered mass-loss models could effectively replicate the bimodal distribution of planet radii in small, close-in exoplanets, as observed by the Kepler space mission. To break this degeneracy, Rogers et al. (2021) introduced a new approach to differentiate between the two distinct models by analyzing the radius valley in the 3D parameter space (period, insolation flux, and stellar mass). Ho & Van Eylen (2023) extended Rogers et al. (2021)'s 3D approach to 4D models by including age as another dimension. Using precise planetary radii based on Kepler 1-min cadence data of 431 planets, they measured slopes in the 2D parameters space for period, stellar mass and insolation flux and suggested that these findings support thermally-driven mass loss models such as photoevaporation and core-powered mass-loss, with a slight preference for the latter scenario. To apply a much larger sample size to the 3D parameter space approach by Rogers et al. (2021), Berger et al. (2023) used uniformly derived stellar and planetary parameters for a sample of 3702 planets. Using this larger sample with slightly larger planetary radius uncertainty, they concluded their results align more closely with core-powered mass-loss than with photoevaporation.

Investigations as to why super-Earths and sub-Neptune are separated by a radius valley at ~1.8 $R_\oplus$ have led to several theoretical models that predict the slope of this valley in log-log space for various parameters such as orbital period, insolation flux and stellar mass. These models range from thermally-driven mass loss processes like XUV photoevaporation and core-powered mass loss, to more evolutionary effects such as gas-poor formation and impact erosion. All four are described below:

- Photoevaporation (PE): In this scenario, intense stellar radiation in the form of extreme ultraviolet and x-rays (XUV) in the first 100 Myr heats the planet's atmosphere, causing it to escape into space. Planets that have lost their atmosphere are expected to have smaller radii and migrate to the super-Earth side of the radius valley.
- Core-Powered Mass-Loss (CPML): This process suggests that planets lose mass due to the energy released by the cooling luminosity of hot rocky cores. As these hot cores cool and dissipate heat over time, they can also heat the planet's atmosphere, causing it to escape which decreases planetary radius, and again migrate to the super-Earth side of the radius valley.
- Gas-Poor Formation (GP): This process hypothesis that planets below the radius valley formed without significant gas envelopes. This would result in a primordial rocky planet population that formed after the dispersal of the protoplanetary disk. This would also result in a large population of super-Earths with smaller radii compared to planets above the valley.
- Impact Erosion (IE): This is where planetesimal impacts can directly strip away the outer layers of a planet's atmosphere upon collision. The impact energy can cause atmospheric particles to be ejected into space, resulting in mass loss. In addition, planets closer to the star may experience more significant mass loss compared to those farther away, which is why this is a potential driver of the radius valley phenomenon.

Other formation processes and models have also been proposed, but the ones that have clear predictions on the slope of the radius valley are summarized in Table 1.

**Table 1:** Theoretical Model Predictions

| Formation Process | Model Predictions (slope) | | | Model Dimensions[a] | Source[b] |
|---|---|---|---|---|---|
| | $d\log R_p/d\log P$ ($R_\oplus$/days) | $d\log R_p/d\log S_p$ ($R_\oplus/S_\oplus$) | $d\log R_p/d\log M_*$ ($R_\oplus/M_\odot$) | | |
| Photo-evaporation | $-0.25 \leq m \leq -0.16$ | | | 2D | (1) |
| | $-0.15$ | $0.11$ | | 2D | (2) |
| | $-0.16$ | | $0.29$ | 3D | (3) |
| Core-Powered Mass Loss | $-0.11$ | | $0.33$ | 2D | (4) |
| | $-0.11$ | | $0.32$ | 3D | (5) |
| Gas Poor Formation | $0.11$ | $-0.08$ | | 2D | (6) |
| Impact Erosion | $-0.33$ | $0.25$ | | 2D | (7) |

[a] 2D & 3D refers to the power law formula used to estimate the slope of radius valley, where 2D is in the form $R_p \propto P^\beta$ and $\beta$ is the slope.
[b] Sources: (1) Owen & Wu (2017), (2) Lopez & Rice (2018), (3) Rogers et al. (2021), (4) Gupta & Schlichting (2020, 2019), (5) Rogers et al. (2021), (6) Lopez & Rice (2018), (7) Wyatt et al. (2020)

To validate any theoretical model and the predicted measurements, observations and associated data analysis must be performed. Studies outlined in Table 2 cover a variety of sample sizes from around 100 to several thousand, with % fractional uncertainties ranging from under 5% to over 10%. In addition, several different analysis methods have been employed to accurately measure the slope of the radius valley.

**Table 2:** Observational Analysis Results

| Planet Sample Size | %$R_p$ Median Uncertainty (%) | Host Stellar Type | Analysis Estimates (slope) | | | Analysis Method[a] | Conclusions | Source[b] |
|---|---|---|---|---|---|---|---|---|
| | | | $d\log R_p/d\log P$ ($R_\oplus$/days) | $d\log R_p/d\log S_p$ ($R_\oplus$/$S_\oplus$) | $d\log R_p/d\log M_*$ ($R_\oplus$/$M_\odot$) | | | |
| 117 | 3.4% | FGK | $-0.09^{+0.02}_{-0.04}$ | | | 2D SVM | Consistent with Photoevaporation. Deep valley | (1) |
| 1633 | 3.7% | FGK | $-0.11^{+0.02}_{-0.02}$ | $0.12^{+0.02}_{-0.02}$ | | 2D Gap Bins | Consistent with Photoevaporation | (2) |
| 2066 | 25.2% | FGK | $-0.319^{+0.088}_{-0.116}$ | | | 2D K-Means | Aligned with Thermally-Driven Mass Loss | (3) |
| 328 | 7.1-9.0% | M | $0.058^{+0.022}_{-0.022}$ | $-0.060^{+0.025}_{-0.025}$ | | 2D Gap Bins | Inconsistent with models. Need more data. | (4) |
| < 3308 | 8.9% | FGKM | | | $0.26^{+0.21}_{-0.16}$ | 2D Gapfit | Aligned with Core-Powered Mass Loss | (5) |
| 27 | 3.3% | M | $-0.11^{+0.05}_{-0.04}$ | | | 3D SVM | Consistent with Photoevaporation | (6) |
| 970 | 5.7% | FGKM | $-0.11^{+0.02}_{-0.02}$ | $0.06^{+0.01}_{-0.01}$ | $0.18^{+0.08}_{-0.07}$ | 2D Gap Bins | Consistent with TDML. Need higher precision data | (7) |
| 121 | ~8% | M | $-0.02^{+0.05}_{-0.05}$ | $0.02^{+0.02}_{-0.02}$ | $0.08^{+0.12}_{-0.12}$ | 2D Gapfit | Planet density, not radii, driver of radius valley | (8) |
| 431 | 4.7% | FGK | $-0.11^{+0.02}_{-0.02}$ | $0.07^{+0.02}_{-0.01}$ | $0.23^{+0.09}_{-0.08}$ | 4D SVM | Deep valley favoring Core-Powered Mass Loss | (9) |
| 3702 | 16.5% | FGKM | $-0.046^{+0.125}_{-0.117}$ | | $0.069^{+0.019}_{-0.023}$ | 3D Gapfit | More consistent with Core-Powered Mass Loss | (10) |

[a]2D, 3D & 4D refers to the power law formula used to estimate the slope of radius valley, where 2D is in the form $R_p \propto P^\beta$ where $\beta$ is the slope. Slope estimation methods are as follows:
SVM refers to Support Vector Machine for gap segmentation
Gap Bins refers to grouping x axis distribution into discrete bins to find position of gap in each bin
K-Means refers to clustering methods to distinguish gap between different clusters
Gapfit refers to gapfit python module (Loyd et al. 2020)
[b]Sources: (1) Van Eylen et al. (2018), (2) Martinez et al. (2019), (3) MacDonald (2019), (4) Cloutier & Menou (2020), (5) Berger et al. (2020), (6) Van Eylen et al. (2021), (7) Petigura et al. (2022), (8) Luque & Pallé (2022), (9) Ho & Van Eylen (2023), (10) Berger et al. (2023)

Most of the research highlighted in Table 2 provided limitations, gaps, and recommendations for further study in better resolving dominant exoplanet formation processes. In general, these gaps and recommendations fell into three categories: 1) the need for more sample planets, 2) the need for more accurate and precise planetary radii which is highly dependent on the host star accuracy and precision, and 3) the need to segment planets according to the mass, metallicity, and age of their host star. The third category is especially important since not only could different dominant formation processes exist for each segment, but individual planets could also have evolved through mechanisms of multiple formation processes. The gaps identified for 1) and 2) can be addressed first and is the focus of this work. Not only can the larger set of planet samples improve our understanding of the planet formation around the radius valley, more precise planetary radii such as the attempts by Fulton et al. (2017) and Martinez et al. (2019) can also illuminate the underlying mechanism previously unnoticeable. In fact, Van Eylen et al. (2018) and Ho & Van Eylen (2023) only used asteroseismic or short cadence transit data to obtain more precise data. However, this greatly limited the sample sizes.

While traditional methods for determining stellar parameters, such as interferometry and spectral analysis, are subject to strict restrictions and biases related to parallax measurements, signal-to-noise ratio spectra, and large blocks of observing time, Spectral Energy Distribution (SED) fitting techniques with proven stellar atmosphere models can provide alternative methods for estimating accurate stellar radii. As shown in Vines & Jenkins (2022), many advantages of SED-based stellar parameter include: (1) a more flexible and inclusive approach, (2) a model-agnostic approach, (3) many stellar parameters obtained simultaneously (temperature, surface gravity, metallicity, extinction, and overall normalization), (4) less sensitive to starspots compared to spectroscopic approaches, and (5) easily applicable to a large number of stars by utilizing commonly available large dataset of broad-band photometric data. By using Bayesian enhanced SED fitting methods, this article focuses on the creation of the largest planet sample with a much better precision in planetary radii estimation that can allow the investigation of the dominant planet formation mechanism around the radius valley.

This article will review the research methods used to create a large high precision sample of planets that reside within the recommended parameters for analyzing the Radius Valley. This will include details on the SED fitting methodology in general to estimate stellar radii, as well as the specific

Bayesian techniques used by SEDmc in this work. We will cover the procedures used to create and maintain a large high-precision sample and discuss the pros and cons of using a variety of methods in estimating key parameters to keep the sample as large as possible. We will then discuss how these parameters are used to determine the slope of the Radius Valley for different power law functions and compare those slopes to theoretical predictions and past observational results. Finally, the estimated slopes obtained in this work will be used to develop conclusions and recommendations for follow-up research.

## 2. Methods

### *2.1. SED Fitting Overview*

The SED fitting process involves comparing the observed SED of a star with synthetic SEDs generated from theoretical stellar atmospheric models. The observed input photometric data used in the SED fitting process can come from many different sources, including large-scale photometric surveys, such as Gaia, PanSTARRs, Sloan Digital Sky Survey (SDSS), the Two Micron All-Sky Survey (2MASS), WISE, etc. There are also several sources of stellar atmosphere models such as the NextGen (Hauschildt et al. 1998) or CK04 (Castelli & Kurucz 2004) model grids. These data sources for input photometry and model grids are used as part of the overall SED fitting process as follows:

1. Observed Data Collection & Filtering: Photometric observations of the star across different wavelengths (e.g., ultraviolet, optical, and infrared) are collected from various telescopes and instruments. When data is collected from different sources, filtering will be needed to ensure fitting is performed using the most consistent and accurate data.
2. Model Selection: Choose the best theoretical stellar atmospheric model(s) that represent a range of stellar parameters such as effective temperature, surface gravity, metallicity, and extinction that are expected from the star being analyzed.
3. Parameter Difference Estimation: Determine the difference between the observed SED and the synthetic SED generated by the models. The parameters typically include effective temperature ($T_{\text{eff}}$), surface gravity (log $g$), metallicity ($[Fe/H]$), extinction ($A_V$), and overall normalization.
4. Fitting Technique: Use statistical methods such as minimizing the $\chi 2$ statistic (Sgro & Song 2021) to find the best-fit parameters that provide the closest match, or maximum likelihood, between the observed and synthetic SEDs.

5. Uncertainty Estimation: Determine the uncertainties associated with the fitted parameters to quantify the precision of the derived stellar properties using Bayesian MCMC sampling and simulation techniques. This enhancement to basic SED fitting (step 4) will be discussed in detail in the next section.
6. Benchmarking and Validation: Compare the results of the SED fitting with other observational methods proven to provide highly accurate results, such as interferometric or asteroseismology measurements of stellar parameters, to validate the accuracy and precision of the derived stellar parameters.
7. Interpretation: Interpret the derived stellar parameters, such as radius, effective temperature, surface gravity, and metallicity, in the context of stellar evolution and astrophysical disciplines that are useful to the field of study requiring stellar parameters, such as estimating accurate and precise exoplanet parameters.

A good example of a detailed implementation of a basic SED fitting process (without step 5) in the Python programming language would be the SED fitting module used in various studies over the past 25 years such as Zuckerman & Song (2004) and Lee et al. (2020). It was the basis of the core SED fitting process used in developing the Python program for this work, called SEDmc, and provided many of the algorithms, functions and data sets used in the Bayesian enhanced SED Fitting code. The methods and results from SEDmc will be discussed in later sections.

## 2.2. Bayesian Enhanced SED Fitting

Adding Bayesian methods to the basic SED fitting process such as the one used in Sgro & Song (2021) enhances SED fitting analyses by incorporating prior knowledge, quantifying uncertainties, addressing biases, and optimizing model parameters. Improving the accuracy and precision in estimating stellar parameters by adding Bayesian methods to the standard SED fitting process has been demonstrated in various research efforts such as analyzing galaxies (Han & Han 2014), active galactic nuclei (AGNs) (Rivera et al. 2016), supernovae (Mandel et al. 2022), as well as estimating general stellar parameters for specific stars (Fabricio Bolutavicius et al. 2023) or a broad group of stars (Vines & Jenkins 2022). The ARIADNE program and the published results (Vines & Jenkins 2022) show that Bayesian methods can be applied to model selection as well as the general SED fitting process and demonstrates the accuracy and precision of this method for a group of 135 interferometrically measured benchmark stars with radii and effective temperatures ranging from 0.15 to 16.57 $R_\odot$ and 2940 to 9377 K respectively. The output of the ARIADNE code compared to benchmark values resulted in a mean fractional difference of 0.001 ±0.070 (1.0%) reported overall. In looking at the output data provided in that study, absolute mean value of fractional residuals for all stars is 0.021 (2.1%) for $T_{\text{eff}}$ and 0.057 (5.7%) for $R_\star$.

For this work, we used a Bayesian enhanced SED Fitting program (SEDmc) we developed for more general stellar parameter estimation research. To verify the accuracy of the SEDmc code, we found 50 unique benchmark stars with values based on either asteroseismology (Heiter et al. 2015) or interferometrically (Karovicova et al. 2020, 2021; White et al. 2018) measured data. The most recent stellar parameter values were used for the 5 duplicate stars from these 5 sources. The results from SEDmc compared to benchmark for $T_{\text{eff}}$ shows the absolute mean value of fractional residuals for all stars is 0.0052 (0.52%). In addition, we derived a median % fractional uncertainty of 1.60% for $T_{\text{eff}}$, compared to the benchmark value of 0.96%. Similarly for the parameter of interest, $R_\star$, Figure 1 shows the absolute mean value of fractional residuals for all stars is 0.0203 (2.03%) and a median % fractional uncertainty of 2.95% compared to the benchmark value of 1.18%. It should be noted that some stars in the benchmark sample, especially those larger than 20 $R_\odot$, are not Main Sequence (MS) stars and may not perform as well as MS stars since the models used were optimized for MS stars.

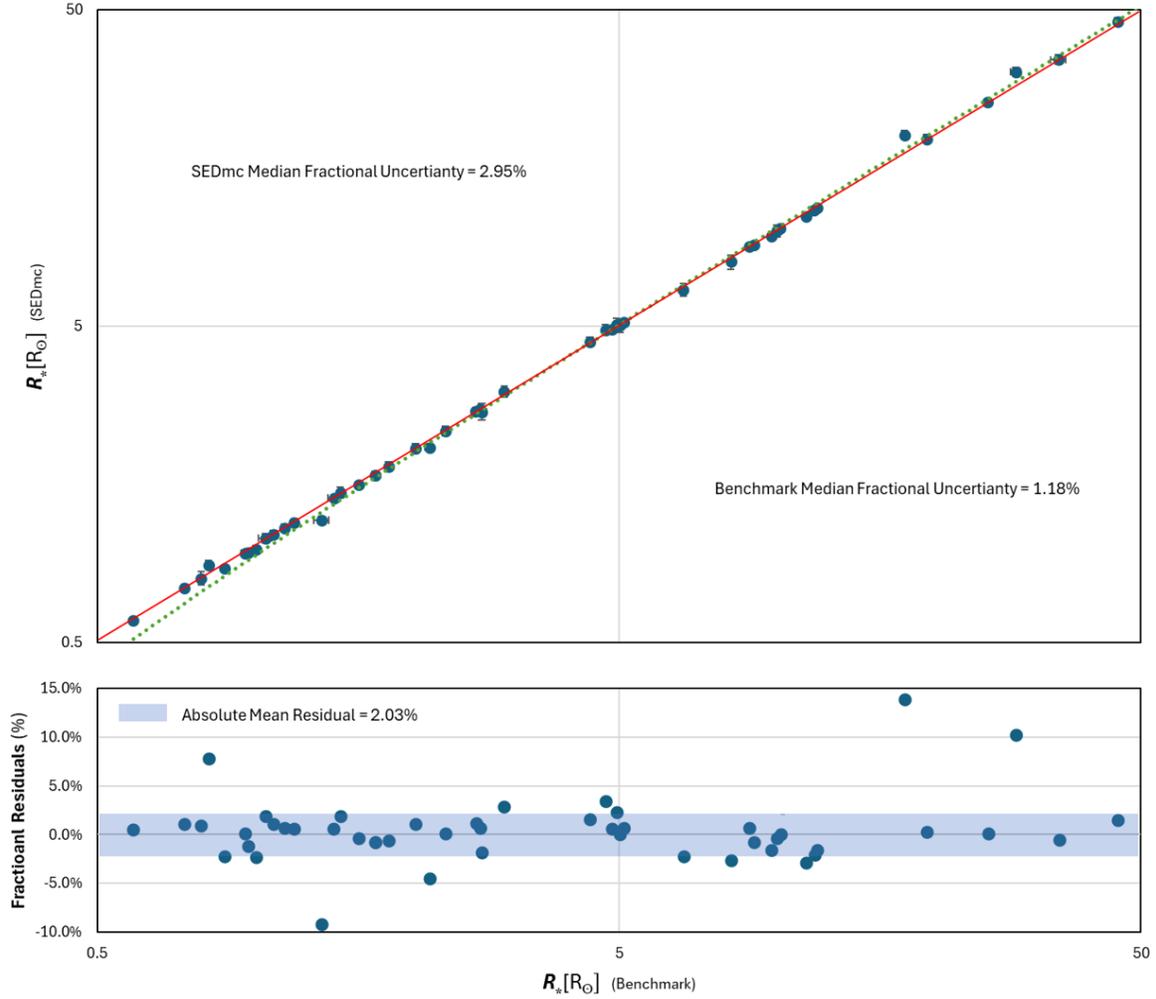

**Figure 1:** The $R_\star$ values and uncertainties estimated with SEDmc compared to those 50 values from benchmark star data. The red line is where they would be equal, and the dotted green line shows the overall trend from low to high $R_\star$. The median % fractional uncertainty of 2.95% is noted as compared to the benchmark value of 1.18%. The lower panel shows fractional residuals for the % difference of SEDmc values from benchmark values where the absolute mean value of fractional residuals for all stars is 2.03%.

### 2.3. SEDmc Methods

SEDmc uses the same process and general methods as SED Fitting with a Markov Chain Monte Carlo (MCMC) sampler package, emcee (Foreman-Mackey et al. 2012), added to improve accuracy and determine precise uncertainties for median values. The Posterior Probability in Bayesian statics is given by Bayes Theorem:

$$P(\theta|D) = \frac{P(D|\theta)P(\theta)}{P(D)} \quad (1)$$

Where $\theta$ is the parameter space of our model, and D is our observed data. The other terms here are:

$P(D|\theta)$: The probability of the data given the model, called the Likelihood.

$P(\theta)$: The probability of our model, known as the Prior.

$P(D)$: The probability of the data, often called the Evidence.

For the emcee sampler, we need to create likelihood and prior functions to determine the posterior probability of each parameter space sample given the observed data. The goal is finding the model parameter values with maximum probability, so we need to define a "maximum likelihood" function. The maximum likelihood function used in our code is:

$$\ln[P(D|\theta)] = -\frac{1}{2}\sum_{i=1}^{N}\frac{(D_i - \theta_i)^2}{\sigma_{D_i}^2} \qquad (2)$$

Instead of trying to define the probability of a particular model parameter $P(\theta)$, with values somewhere between 0 and 1, we can just use it to filter the sampler moves (i.e., sampler randomly selected values) within a range of valid parameter values. Therefore, our "prior" knowledge of the parameter space only needs to be the valid range of parameter values. If the randomly selected model parameter value is within the range specified, the prior probability is equal to 1 and the posterior probability is just equal to the likelihood probability. If it is not within that range, then the posterior probability is equal to 0, which effectively eliminates that sample from the final distribution. The exact details of how this is done can be found in the emcee documentation (Foreman-Mackey et al. 2012).

The primary enhancement to the SED code is the ability to determine statistically valid uncertainties of the estimated parameter values. The emcee sampler, if successful, will provide a pseudo-normal probability distribution for each parameter. The Python code for Bayesian enhanced SEDmc module and other tools used in this work can be found at GitHub[3].

### 2.4. Sample Selection & Filtering

To create a large high-precision sample of planet radii and insolation flux to be used in analyzing the radius valley, we start with the most current exoplanet data from NASA Exoplanet Archive site (NASA Exoplanet Archive, 2024). An original sample of 5638 confirmed exoplanets with planetary and host stellar data was downloaded from the NASA site on 6/2/2024. Only confirmed planets were selected to ensure the integrity of the radius valley analysis in focusing on verified planet data. Seeing as SEDmc will provide high precision stellar data such as $R_\star$ and $T_{\text{eff}}$, the goal was to derive planetary parameters from the highly accurate and precise stellar parameters.

---

[3] https://github.com/proxcent/Radius_Valley

Given this goal, the original sample was filtered down according to the availability of other data needed to make those calculations, including transit data to calculate planet radii $R_p$ and orbital data to calculate insolation flux $S_p$. In addition, the radius valley formation analysis is only relevant to orbital periods between 1 and 100 days and planet radii between 1 and 4 $R_\oplus$. Thus, the sample had to be filtered further to support a valid analysis of the valley. Finally, to ensure the radius valley slope estimation is as accurate as possible, only planet radii with % fractional uncertainty less than 5.0% was used as recommended by Rogers et al. (2021). The complete filtering pipeline is shown in Table 3, which resulted in 1434 highly accurate and precise host stars values that were then used to derive highly accurate and precise planetary parameters for 1923 exoplanets distributed within the boundaries of the radius valley. The next sections will describe those results.

**Table 3**: Radius Valley Exoplanet Sample Filtering Pipeline

| Filter Criteria | Hosts | Planets | Description |
|---|---|---|---|
| Original Sample | 4199 | 5638 | Exported from NASA Exoplanet Archives on 6/2/2024 |
| Transit & Parallax Data Available | 2981 | 3949 | Needed for planet radius $R_p$ calculation |
| Orbit Semi-Major Axis Data Available | 2847 | 3771 | Needed for Insolation Flux $S_p$ calculation |
| Filter for Period > 1 & < 100 days | 2664 | 3499 | Scope of Radius Valley Analysis |
| Filter for $R_p$ > 1 & < 4 $R_\oplus$ | 1873 | 2538 | Scope of Radius Valley Analysis |
| SEDmc Results for 1873 Host Stars | 1873 | 2538 | SEDmc had 100% success rate in estimating $R_\star$ |
| Re-Filter for $R_p$ > 1 & < 4 $R_\oplus$ | 1803 | 2443 | Scope of Radius Valley Analysis |
| $R_p$ %Fractional Uncertainty < 5.0% | 1434 | 1923 | Final Sample - Need high precision $R_p$ data for good estimate of Radius Valley slope |

Using the batch processing function of SEDmc, 1873 hosts stars were analyzed using Bayesian enhanced SED fitting which provided successful results and stellar parameters as described previously. Since SEDmc uses two different sources of stellar atmosphere models including the NextGen (Hauschildt et al.

1998) and CK04 (Castelli & Kurucz 2004) model grids, both were used for each star and the data with the highest precision was used. The CK04 model doesn't contain models for stars with $T_{eff}$ less than 3500 K. Therefore, the NextGen model was used for any star where the $T_{eff}$ in the NASA data was under 3500 K. Once we filtered for $R_p$ % fractional uncertainty less than 5.0%, the 1434 remaining host stars had a median % fractional uncertainty of 2.4% as shown in Figure 2. We also note that median % fractional uncertainty from the NASA Archive for 1394 of these same stars that had error data is 9.1%. Figure 2 also shows how the different models performed in estimating $R_\star$ as compared to the NASA data, along with 39 stars with asteroseismic (AS) derived values from Van Eylen et al. (2018). The trends shown by the linear fit lines for each model do not indicate strong biases for either model with CKO4 trending slightly larger for smaller stars. However, in comparing the distribution statistics for these results we see CK04 has a lower mean and variance, with the t-Test statistics showing a notable difference. The mean residual for all stars is 6.0%, and 3.2% for AS derived stars. Two of the AS derived stars (Kepler-1002 and Kepler-1392) have much higher residuals than the rest. If those two outliers are removed, the mean residual for AS derived stars is 2.2% showing SEDmc can provide high accuracy estimates for the stellar radius for this sample.

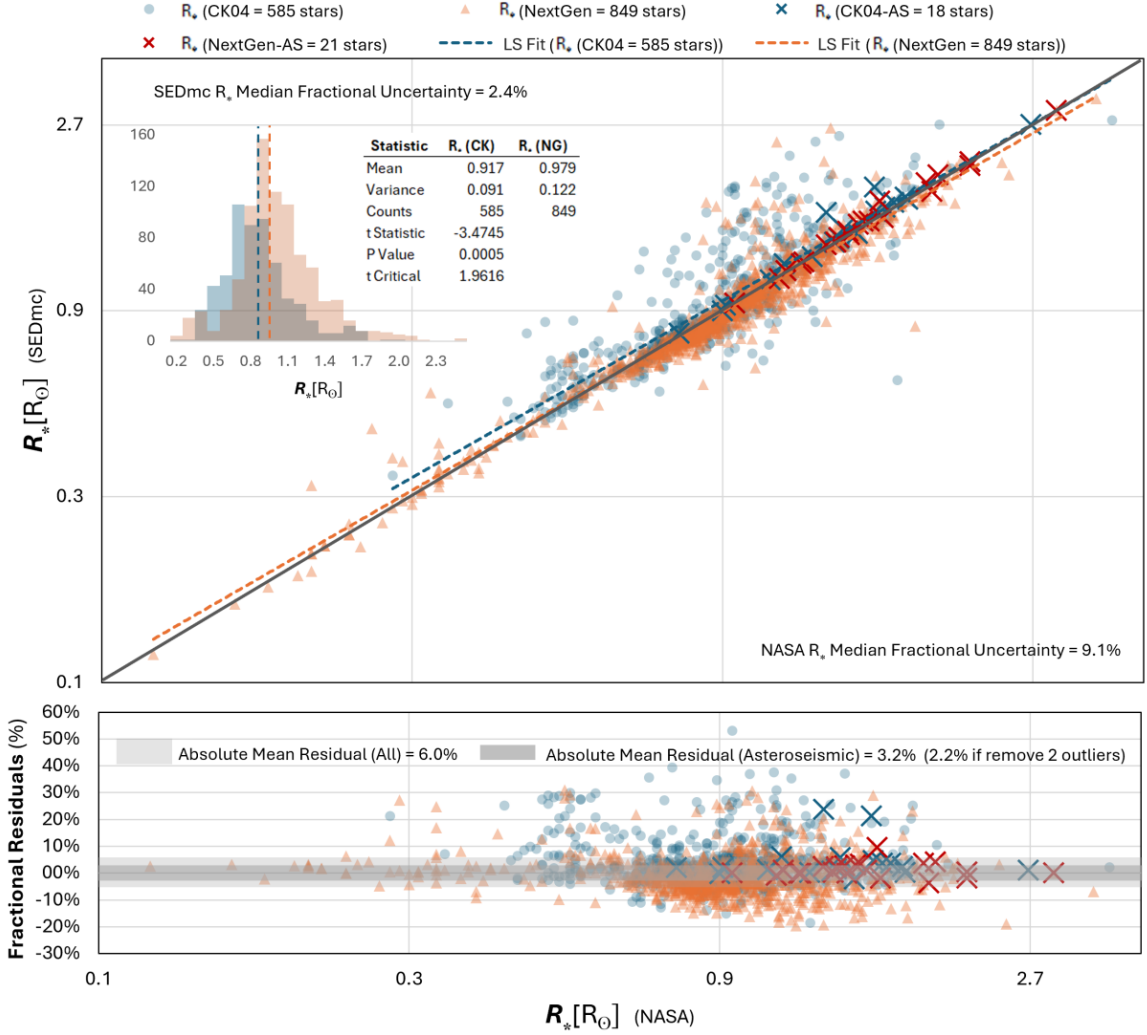

**Figure 2:** The 1434 $R_*$ values estimated with SEDmc compared to the values from the NASA Archive data. The legend shows the markers used for SEDmc estimates by model where the blue circles used the CK04 model, and the orange triangles used the NextGen model. Upper: The diagonal line shows where they would be equal, and the dotted lines show the linear fit from low to high $R_*$, where blue is for CK04, and red is for NextGen data. The X markers are for 39 stars with asteroseismic (AS) derived values from Van Eylen et al. (2018). The median % fractional uncertainty for the SEDmc derived data is 2.4% compared to that of the NASA data of 9.1%. Model distribution statistics are shown. Lower: Fractional residuals for the % difference of SEDmc values from NASA values where the absolute mean value of fractional residuals for all stars is 6.0%. When comparing SEDmc to the AS derived values, the fractional residuals are reduced to 3.2% for all 39 stars, and 2.2% for 37 stars.

### 2.5. Planet Radius Calculations

With the stellar radii % fractional uncertainties well under 3% in most cases, we used the transit depth provided by the NASA Exoplanet Archive to calculate accurate and precise planet radii. We used this to calculate the radius of the planet using the relationship,

$$R_p = 109.27 R_* \sqrt{\delta_t}, \qquad (3)$$

Where $\delta_t$ is the transit depth in ratio of star brightness blocked by the planet and the constant 109.27 is the ratio of the sun's radius to the earth radius so $R_p$ is in units of $R_\oplus$ when $R_\star$ is in units of $R_\odot$. Another way to calculate planet radius is by using the ratio of planet to star radius provided by the NASA Exoplanet archive $(R_p/R_\star)_{NASA}$. Using this value, which is typically derived from transit depth data similar to equation 3, we also used the following relationship,

$$R_p = 109.27(\frac{R_p}{R_*})_{NASA}, \qquad (4)$$

Usually, the method that should result in the highest precision would be the more direct calculation of equation 3. However, inconsistencies in the NASA reported data resulted in some planets with higher precision using equation 4. Therefore, to keep the sample as large and precise as possible, we chose the method that provided the highest precision for planet radius once uncertainties were propagated. To determine the impact of this method we show the $R_p$ vs. Period plot in the left panel of Figure 3 for both equations, along with distribution statistics. This shows the ratio method has a smaller mean and larger variance, and the t-Test indicates a notable difference in those samples.

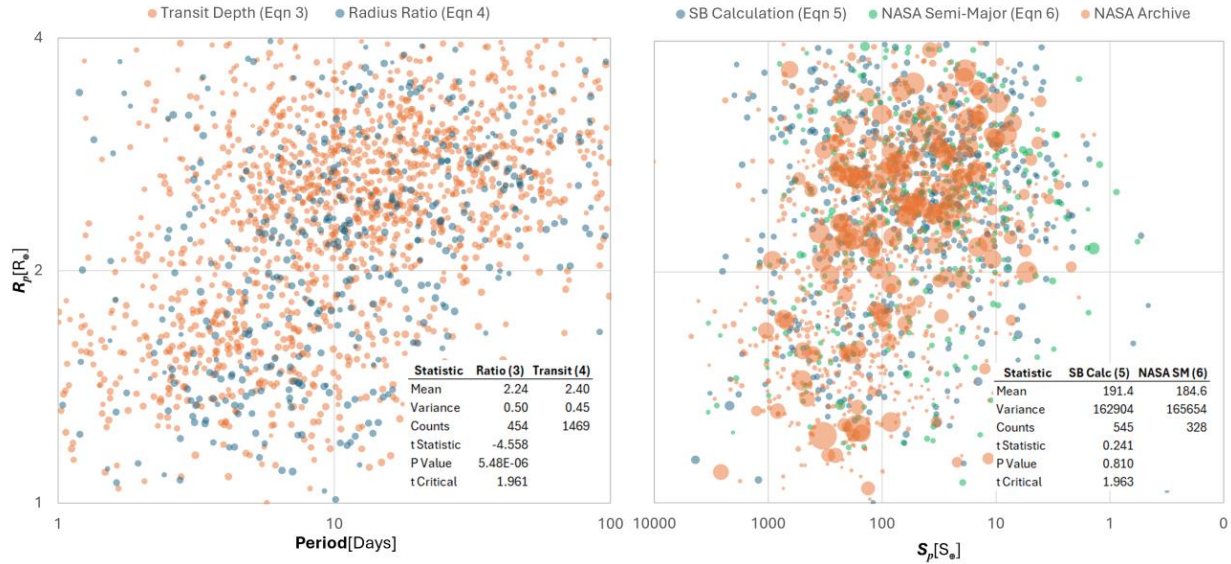

**Figure 3**: Left: Plot of $R_p$ vs. Period for all 1923 planets with the equations used in the $R_p$ calculation shown by different colored markers. Right: Plot of $R_p$ vs. $S_p$ for all planets with the equations used in the $S_p$ calculation shown by the different colored markers. The size of the markers represents uncertainty. Distribution statistics for each equation are shown in the inset table.

This method for calculating $R_p$ provided a median % fractional uncertainty of 3.4% for the 1923 planets shown in Figure 4. In contrast, the median % fractional uncertainty provided by NASA for 1744 of these same planets that had error data is 10.4%. These high precision values of planet radii further confirm and highlight the presence of the radius valley as shown in Figure 4. A clear gap and division between the generally accepted radii (Sotzen et al. 2021) of super-Earths and sub-Neptunes is visible in the chart.

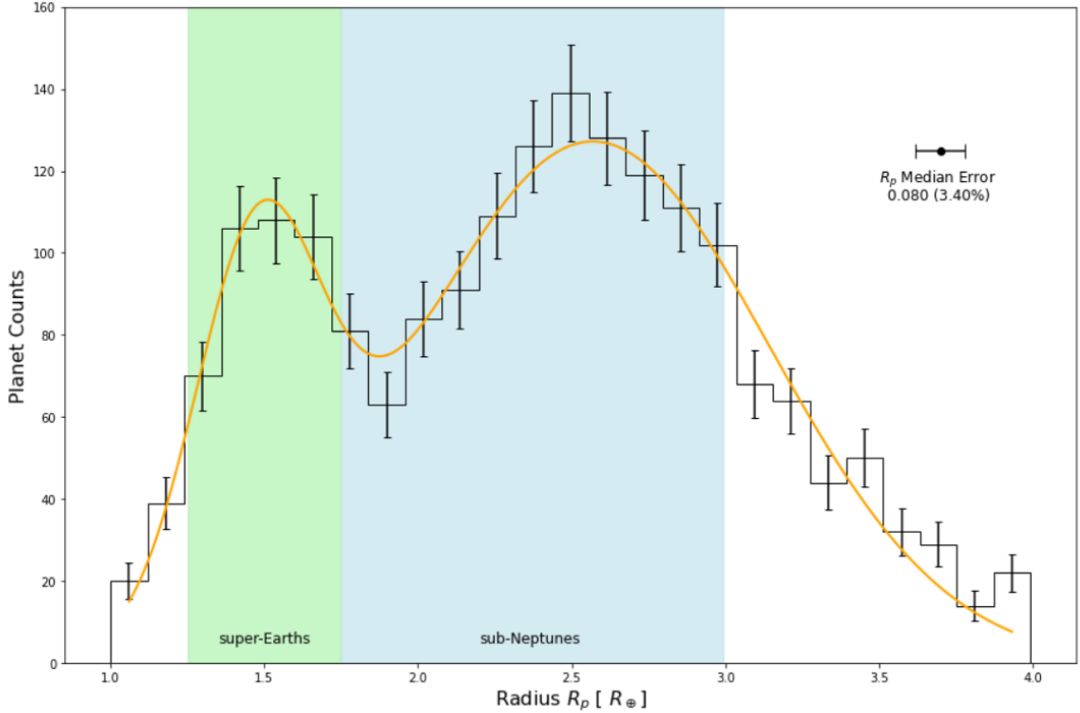

**Figure 4**: Distribution of planet radii of 1923 planets used for the radius valley analysis with 1σ errors and the % median error for $R_p$ of 3.4%. Planet classifications are shown by colored regions as defined in Sotzen et al (2021) highlighting that the radius valley is generally the separation between super-Earths and sub-Neptunes.

### 2.6. Planet Insolation Flux Calculations

To use the model predictions shown in Table 1 to determine the planet formation mechanism for the radius valley we will need to estimate the slope of the valley for radius vs. insolation flux also. For insolation flux we used the high precision stellar parameters $R_\star$ and $T_{eff}$ to estimate a high-precision luminosity for the star using the Stephan-Bolzman Law relationship. Once we know $L_\star$, we can use that, along with the semi-major axis information from the NASA Exoplanet Archive, to estimate the insolation flux $S_p$ for a planet using the following relationship,

$$\frac{S_p}{S_\oplus} = \left(\frac{L_\star}{L_\odot}\right)\left(\frac{AU}{a}\right)^2, \tag{5}$$

where $S_p$ is in terms of Earth's value, $L_\star$ is in terms of its solar value and the semi-major axis $a$ is in AU. We note that we assumed a circular orbit in Equation 5. Planets that are relatively small and close to their host star will generally have circular orbits since the center of mass of these two bodies will be inside the host star. Instead of using semi-major axis data from NASA to calculate planet insolation flux, we could also derive this parameter for planetary orbits using the following relationship,

$$a = \left(\frac{P^2 G M_*}{4\pi^2}\right)^{1/3}, \tag{6}$$

where $P$ is the orbital period, $G$ is the Gravitational constant and $M_*$ is the mass of the host star. Using equation 6 is only desirable when the fractional uncertainties for $P$ and $M_*$ provided by the NASA archive is much less than those provided for the semi-major axis $a$, resulting in a lower fractional uncertainty for insolation flux. Therefore, to keep the sample as large and precise as possible, we chose the method that provided the highest precision for insolation flux once uncertainties were propagated. To determine the impact of this method we show the $R_p$ vs. $S_p$ plot in the right panel of Figure 3 for both equations 5 and 6, as well as the values directly from the NASA archive, along with distribution statistics. This shows similar mean and variance and passed the t-Test with a P value of 0.81, which indicates no notable difference in those samples. Using this method, along with equations 5, and 6, we estimated $S_p$ for the planet sample with a mean % fractional uncertainty of 4.9% which was an improvement over the NASA data of 5.5% for this sample of planets.

### 2.7. Improved Accuracy and Precision of Final Sample

A snapshot of 10 of the 1923 planets in the final sample used for the radius valley formation analysis described in the following sections is shown in Table 4 and the entire list can be downloaded in the online version of this article and on [GitHub](GitHub). This full list shows the planetary and stellar parameters, including uncertainties, for orbital period, planet radius, insolation flux and stellar mass. Values and uncertainties for orbital period and stellar mass were taken directly from the NASA Exoplanet Archive, whereas improved values and uncertainties for planet radius and insolation flux were derived from the high precision stellar temperature and radius from SEDmc as described above.

**Table 4:** List of 10 of the 1923 planets in final sample showing planetary and stellar parameters.

| Planet | Orbital Period[a] $P$ (days) | Planet Radius[b] $R_p$ ($R_\oplus$) | Planet Insolation Flux[c] $S_p$ ($S_\oplus$) | Stellar Mass[a] $M_*$ ($M_\odot$) |
|---|---|---|---|---|
| TOI-411 b | $4.04030^{+0.00002}_{-0.00001}$ | $1.357^{+0.050}_{-0.050}$ | $493.54^{+24.25}_{-25.76}$ | $1.10^{+0.04}_{-0.04}$ |
| TOI-411 c | $9.57308^{+0.00001}_{-0.00001}$ | $2.319^{+0.063}_{-0.064}$ | $156.51^{+7.64}_{-8.16}$ | $1.10^{+0.04}_{-0.04}$ |
| HD 73583 b | $6.39804^{+0.00001}_{-0.00001}$ | $2.887^{+0.083}_{-0.089}$ | $43.73^{+1.96}_{-2.04}$ | $0.73^{+0.02}_{-0.02}$ |
| HD 73583 c | $18.87974^{+0.00086}_{-0.00074}$ | $2.543^{+0.087}_{-0.087}$ | $10.33^{+0.46}_{-0.48}$ | $0.73^{+0.02}_{-0.02}$ |
| GJ 1132 b | $1.62893^{+0.00003}_{-0.00003}$ | $1.063^{+0.042}_{-0.043}$ | $19.55^{+1.49}_{-1.53}$ | $0.18^{+0.02}_{-0.02}$ |
| WASP-47 d | $9.03050^{+0.00001}_{-0.00001}$ | $3.555^{+0.108}_{-0.115}$ | $144.65^{+9.72}_{-10.12}$ | $1.06^{+0.05}_{-0.05}$ |
| TOI-815 b | $11.19726^{+0.00002}_{-0.00002}$ | $2.853^{+0.071}_{-0.072}$ | $35.26^{+1.66}_{-1.67}$ | $0.78^{+0.04}_{-0.04}$ |
| TOI-815 c | $34.97615^{+0.00010}_{-0.00010}$ | $2.550^{+0.099}_{-0.107}$ | $7.72^{+0.36}_{-0.37}$ | $0.78^{+0.04}_{-0.04}$ |
| TOI-1235 b | $3.44471^{+0.00009}_{-0.00009}$ | $1.721^{+0.074}_{-0.079}$ | $57.33^{+2.90}_{-3.12}$ | $0.63^{+0.02}_{-0.02}$ |
| TOI-733 b | $4.88477^{+0.00002}_{-0.00002}$ | $1.957^{+0.098}_{-0.093}$ | $251.40^{+12.10}_{-15.18}$ | $0.96^{+0.05}_{-0.03}$ |

[a]Taken directly from NASA Exoplanet Archive
[b]Calculated using equations 3 and 4
[c]Calculated using equations 5 and 6 or taken directly from the NASA Exoplanet Archive
(This table is available in its entirety in machine-readable form.)

By filtering the original sample to obtain the right transit and orbital data needed to calculate the higher precision planetary parameters using the results of the Bayesian enhanced SEDmc code, we were able to improve planet radii precision over all other previous studies for a sample size greater than 1000 planets. A comparison of this work with previous studies is shown in Figure 5 with both sample size and median % fractional uncertainty of $R_p$ displayed as vertical bars. The sample size for this work is shown as a blue bar indicating 1923 planets, with the % fractional uncertainty of $R_p$ as an orange bar indicating 3.4%. The only previous study that had a smaller median % fractional uncertainty just had 27 planets in the sample (Van Eylen et al. 2021). This demonstrates that this work was able to achieve the largest sample size with the highest precision to date.

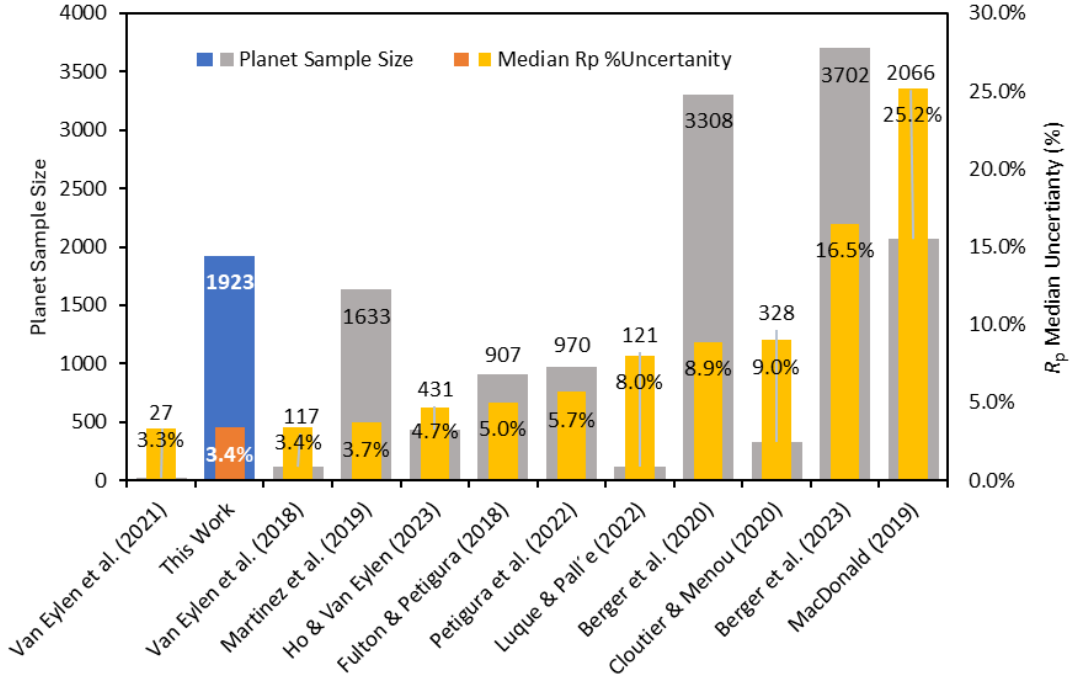

**Figure 5:** This work compared to previous studies. The sample size for this work is shown as a blue bar indicating 1923 planets, with the % fractional uncertainty of $R_p$ as an orange bar indicating 3.4%.

## 2.8. Radius Valley Slope Measurement

Planet formation analysis was performed using the high precision sample of 1923 planets obtained through selective filtering and Bayesian SED fitting. This required various analysis techniques so the results of this sample can be compared with the planet formation models summarized in Table 1, as well as previous observational analyses outlined in Table 2. Typically, we try to identify trends in where data is abundant, but in this case our goal is to measure a trend in the absence of data (Rogers et al. 2021). As the various models indicate, the slope of the radius valley for radius vs. orbital period and radius vs. insolation flux are important metrics for testing theoretical predictions. However, characterizing the absence of planets makes fitting a parameterized description of the radius valley challenging (Petigura et al. 2022). Given this challenge, we have investigated multiple ways to measure the slope as a function of planetary and stellar parameters shown in Table 4. These also include visualization aids such as Kernal Density Estimation (KDE) plots that can highlight the density of planet populations and emphasize the area and contours of the radius valley where planets are sparse.

Several techniques for measuring the slope of the radius valley have been used in previous studies with varying degrees of success. These techniques have evolved over the last few years from the 2D method used by most, to the more recent 3D (Berger et al. 2023; Van Eylen et al. 2021; Rogers et al. 2021) and 4D (Ho & Van Eylen 2023) techniques which can be somewhat sophisticated and challenging to implement. Other methods and tools used include what we are calling "Gap Bin" (Cloutier & Menou 2020; Martinez et al. 2019; Petigura et al. 2022), Support Vector Machines (SVM) (Van Eylen et al. 2018, 2021; Ho & Van Eylen 2023) and using the gapfit Python module (Berger et al. 2020, 2023; Loyd et al. 2020; Luque & Pallé 2022) developed by Loyd et al. (2020).

The Gap Bin method is a piecewise segmentation that creates distinct bi-modal distributions for measuring points along the radius valley. This method involves segmenting the data for the horizontal axis, whether orbital period, insolation flux or stellar mass, into distinct groups. We used a 2D power law formula $R_p = X^m e^{R_{p0}}$ and estimated the slope $m$ and $R_{p0}$ intercept using the Gap Bin slope measurement process with additional filtering and statistical methods to reduce computational biases and determine uncertainties. Specifically, $X$ could be either horizontal parameter $P$ (orbital period), $S_p$ (planet insolation flux) or (stellar mass) $M_\star$, where $R_{p0}$ is the value of $R_p$ where the line intercepts the horizontal axis at $X = 0$. The analysis process for all planetary and stellar parameters is as follows:

1. Convert all data to log-log values so the slope can be estimated and visualized in log-log scale.
2. Use the Gap Bin method for group sizes ranging from 1 to 20 and for bin sizes ranging from 10 to 60. In theory, this can generate over 3000 slope data points, but there will be group and bin size combinations where bi-modal gaussian curves cannot be generated. In general, there will usually be 100 to 1000 data points available for statistical analysis.
3. Remove data outliers using interquartile range (IQR) filtering where we calculate the $IQR$ value. We then take 1.5 times the $IQR$ and subtract this value from 1st quartile (Q1) and add this value to the 3rd quartile (Q3) to establish outlier thresholds. Any data points less than 1.5$IQR$ below Q1 or more than 1.5$IQR$ above Q3 are considered outliers and discarded.
4. Use linear regression as a sanity check to get an initial slope and intercept which can be used as an initial guess for step 5.
5. Then use Bayesian methods to estimate the final slope and the associated uncertainties. Bayesian methods are preferred due to the proven accuracy of Affine Invariant Markov Chain Monte Carlo (MCMC) Ensemble sampling used with emcee Python module.

6. Overlay the results of the Bayesian method to determine the slope and intercept for the 2D power law equation on the KDE contour plot of the sample planets for each horizontal parameter (i.e., orbital period, insolation flux or stellar mass).
7. Repeat steps 2 through 6 using a filtered sample of stellar masses between 0.8 and 1.2 $M_\odot$ to see if planets with host stars near 1 $M_\odot$ show better alignment to model data or previous observational studies. We will present these findings as additional data to consider for the final conclusions.
8. Show additional visualizations when comparing the results to model data or previous observational studies to help clarify the conclusions expressed in this work. Summary plots and visualizations will be shown in the Results and Discussion section.

As shown in previous studies using this method, the results are sensitive to the sample size, accuracy and precision. This dependency will become more evident when we compare the results of the full sample with that of the filtered sample by stellar mass. We also noticed sensitivity when slightly increasing the sample size by 2% through increasing the precision filtering by 0.1%. This sensitivity highlights the limitations of the 2D methods in distinguishing between the two thermally-driven mass loss processes as outline previously (Rogers et al. 2021).

## 3. Results & Discussion

### 3.1. Radius Valley Slope Estimation Results

After filtering out cases with $R_p$ % fractional uncertainty less than 5.0%, the 1434 remaining host stars had a median % fractional uncertainty of 2.4% and the resultant sample of 1923 planets had a median % fractional uncertainty of 3.4% as shown in Figure 4. To estimate the slope of the radius valley for the planet radius vs. orbital period using the 2D power law formula $R_p = P^m e^{R_{p0}}$, the planet radii and orbital period are first converted to log10 values. Next, we use the Gap Bin method described previously and generated slope data points representing various combinations of group and bin sizes to eliminate the bias associated with those input parameters. These slope data points were further analyzed using Bayesian methods of the emcee module that verified the actual slope $m$ and the $R_{p0}$ intercept while also providing uncertainties. The resultant values to 3 decimal places in the power law equation for the slope of the radius valley for the planet radius vs. orbital period for the full sample of 1923 planets is $R_p = P^{-0.142^{+0.006}_{-0.006}} e^{0.896^{+0.012}_{-0.010}}$.

We also performed the slope analysis for a reduced sample of 1292 planets with host star stellar masses between 0.8 and 1.2 $M_\odot$ as mentioned in step 7 of the slope estimation analysis procedure. This was recommended for estimating the slope for insolation flux (Rogers et al. 2021) so we used the same filtered sample for all slope estimations. Since incident insolation flux is strongly correlated with stellar mass, the radius valley position and slope can be influenced by the stellar mass distribution. Shifting the stellar mass distribution to lower masses can reduce certain features in the data, leading to improved accuracy. The resultant values to 3 decimal places in the power law equation for the slope of the radius valley for the planet radius vs. orbital period for the filtered sample of 1292 planets is $\boldsymbol{R_p = P^{-0.097^{+0.002}_{-0.002}} e^{0.716^{+0.005}_{-0.005}}}$.

The plots for planet radius vs. orbital period are shown in Figure 6, where the left panel shows these results compared to theoretical model predictions and the right panel shows the same results compared to past observational analyses. In looking at the left panel, both the results for the full sample of 1923 planets, as well as the sample of 1292 planets filtered by mass are shown along with their shaded uncertainty regions. Additional plots of the 7 different theorical model predictions are also shown, with two of those being the same line for core-powered mass loss (Gupta & Schlichting 2020, 2019; Rogers et al. 2021). The best alignment for the full sample (green line) is the photoevaporation prediction from (Lopez & Rice 2018). However, we see that the filtered sample (blue line) is closer to core-powered mass loss line. Given the analysis for both samples are good estimations based on that sample, this indicates that primary formation process is a thermally-driven mass loss process, but the specific process, whether PE or CPML depends on the choice of stellar mass segmentation for the radius vs. orbital period analysis. More stellar mass segmentation analysis needs to be done to confirm this correlation.

The right panel of Figure 6 again shows planet radius vs. orbital period where both the results for the full sample and planets filtered by mass are shown along with their uncertainties. Additional plots of the 9 different previous observational studies are also shown. Four of those (Van Eylen et al. 2021; Ho & Van Eylen 2023; Martinez et al. 2019; Petigura et al. 2022) had the same slope which is shown as black dashed line. The best alignment with the full sample is those 4 results. However, the filtered sample is in between that line and the line from Van Eylen et al. (2018), but is closer to the latter. Those 4 studies concluded the primary driver is a TDML process with two leaning more to PE and one leaning more to CPML. Therefore, the conclusion of this study also indicates the primary formation process is a thermally-driven mass loss process, but the specific process, whether PE or CPML, may become clearer when comparing our results for insolation flux.

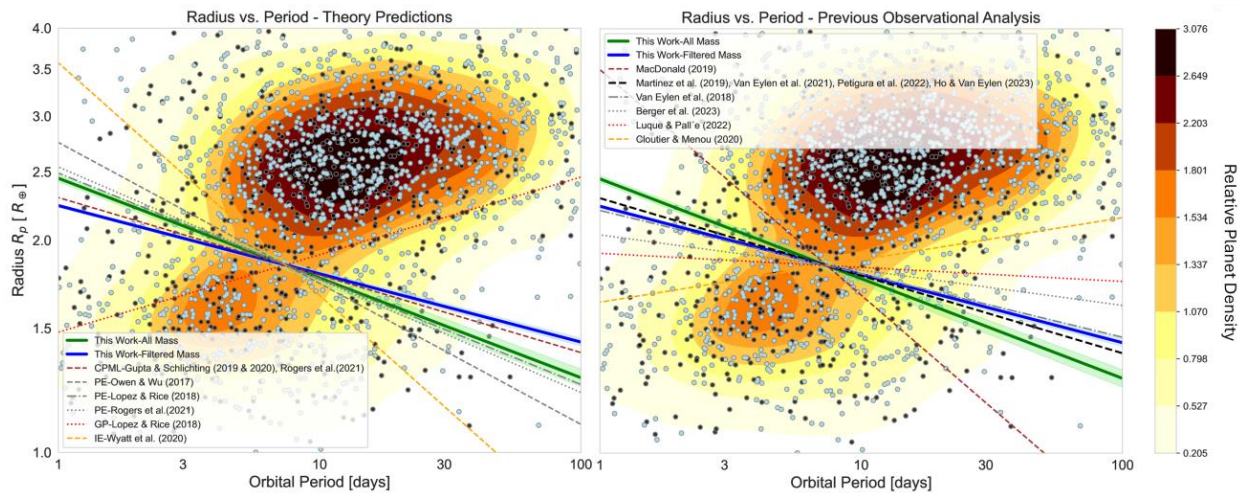

**Figure 6**: Plots for the slopes of planet radius vs. orbital period for full sample of 1923 planets (green), as well as the sample of 1292 planets (blue) filtered by mass are shown along with their shaded uncertainty regions. The KDE plot is for the full sample with light blue markers showing the planets for the filtered mass. Left: Additional plots of the 7 different theoretical model predictions are also shown. Right: Additional plots of the 9 different previous observational studies are also shown. Four of those had the same slope which is shown as black dashed line.

Similarly, the power law equation for the slope of the radius valley for the planet radius vs. insolation flux for the full sample of 1923 planets is $R_p = S_p^{0.041^{+0.001}_{-0.001}} e^{0.366^{+0.002}_{-0.002}}$. The resultant slope and position for the filtered sample of 1292 planets is $R_p = S_p^{0.136^{+0.014}_{-0.014}} e^{-0.085^{+0.031}_{-0.030}}$.

Plots for planet radius vs. planet insolation flux are shown in Figure 7 including 3 different theoretical model predictions. Neither of our samples aligns with any of the 3 model predictions but both are closest, especially the filtered sample, to the TDML process of PE (Lopez & Rice 2018). The slope seems to be more dependent on stellar mass segmentation than the orbital period analysis, but more segmentation would be needed to verify that initial conclusion. Our filtered sample result is close to Martinez et al. (2019) who conclude the PE process. The full sample is closest to Petigura et al. (2022) who concluded it was a TDML process. Even though the filtered sample leans toward a PE process, we feel the $R_p$ vs. $S_p$ slope analysis only further confirms some TDML process is dominant.

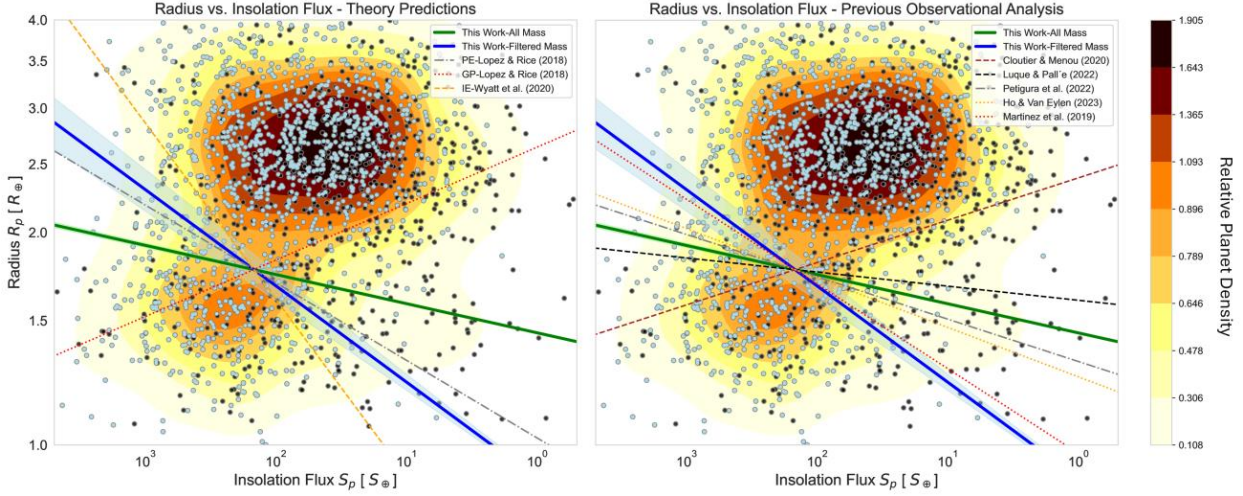

**Figure 7:** Plots for the slopes of planet radius vs. planet insolation flux for full sample of 1923 planets (green), as well as the sample of 1292 planets (blue) filtered by mass are shown along with their shaded uncertainty regions. The KDE plot is for the full sample with light blue markers showing the planets for the filtered mass. Left: Additional plots of the 3 different theoretical model predictions are also shown for PE, GP, and IE. Right: Additional plots of the 5 different previous observational studies.

For radius vs. stellar mass, the slope and position for the full sample is $R_p = M_*^{0.046^{+0.003}_{-0.003}} e^{0.61^{+0.003}_{-0.003}}$ and $R_p = M_*^{0.090^{+0.026}_{-0.025}} e^{0.625^{+0.003}_{-0.003}}$ for the filtered sample. Similarly, we show the plots for planet radius vs. stellar mass in Figure 8. The slope didn't change significantly for the filtered sample. Neither of our samples align well with any of the 3 model predictions. On the right panel of Figure 8, the result from the filtered sample shows a close alignment to Berger et al. (2023) and Luque & Pallé (2022). Even though this is not consistent with model predictions, Berger et al. (2023) concluded their findings were an indicator of core-powered mass loss but cautioned that slope estimation techniques can produce systematic offsets to these results. For the $R_p$ vs. $M_*$ slope analysis, no firm conclusion can be determined but there is some good alignment with TDML with CPML as a possibility.

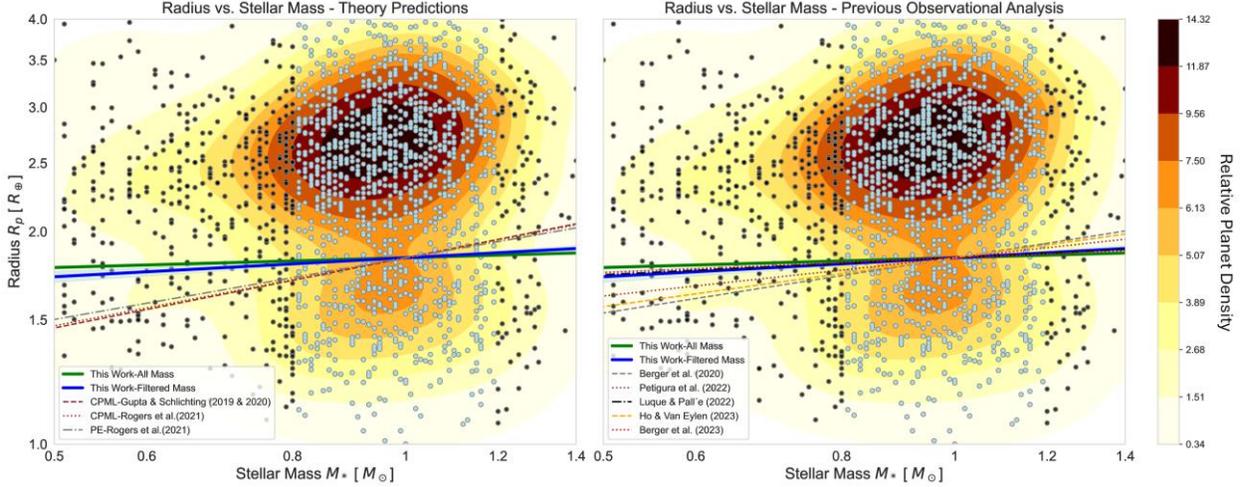

**Figure 8**: Plots for the slopes of planet radius vs. stellar mass for the full sample of 1923 planets (green), as well as the sample of 1292 planets (blue) filtered by mass are shown along with their shaded uncertainties regions. The KDE plot is for the full sample with light blue markers showing the planets for the filtered mass. Left: Additional plots of the 3 different theoretical model predictions are also shown for CPML and PE. The slope didn't change significantly for the filtered sample. Right: Additional plots of the 5 different previous observational studies are also shown.

### 3.2. Combined Model Prediction & Observational Research Results

In addition to viewing all the slopes on the same plot to determine alignment, another way to compare the results to both model predictions and previous observational results is to visualize the relative positions of these results with all those data points on the same diagram. Figures 9 and 10 are a way to better understand those correlations and make additional inferences on how all this data correlates to the planet formation models. This view further confirms the correlation with TDML predicted by the models from measuring the slope as a function of orbital period. Even though these results don't exactly align with model predictions, they are all within the range for TDML. This is evident on the horizontal axis (slope as a function of orbital period) in both figures with the full sample aligning more with photoevaporation model predictions (Lopez & Rice 2018) in Figure 9 and the filtered sample aligning more with CPML (Gupta & Schlichting 2020, 2019; Rogers et al. 2021) in Figure 10. The filtered sample has additional correlation to the observed results of Petigura et al. (2022), Martinez et al. (2019) and Ho & Van Eylen (2023) all confirming TDML and some leaning toward CPML as shown in Figure 9.

Also, the results from measuring the slope as a function of insolation flux on the vertical axis of Figure 9 are relatively close to the results of Martinez et al. (2019) who concluded PE, while the filtered sample is closer to Petigura et al. (2022) and Luque & Pallé (2022) who didn't have firm conclusions other than TDML. Even though Figure 10 does not show clear alignment with model predictions as a function of stellar mass on the vertical axis, both the full sample and the filtered sample is relatively close to Berger et al. (2023) who concluded their findings was an indicator of core-powered mass loss. However, the uncertainties for both were large as shown in the figures. In fact, the results for the slope as a function of stellar mass in most of the past observational studies had large uncertainties also.

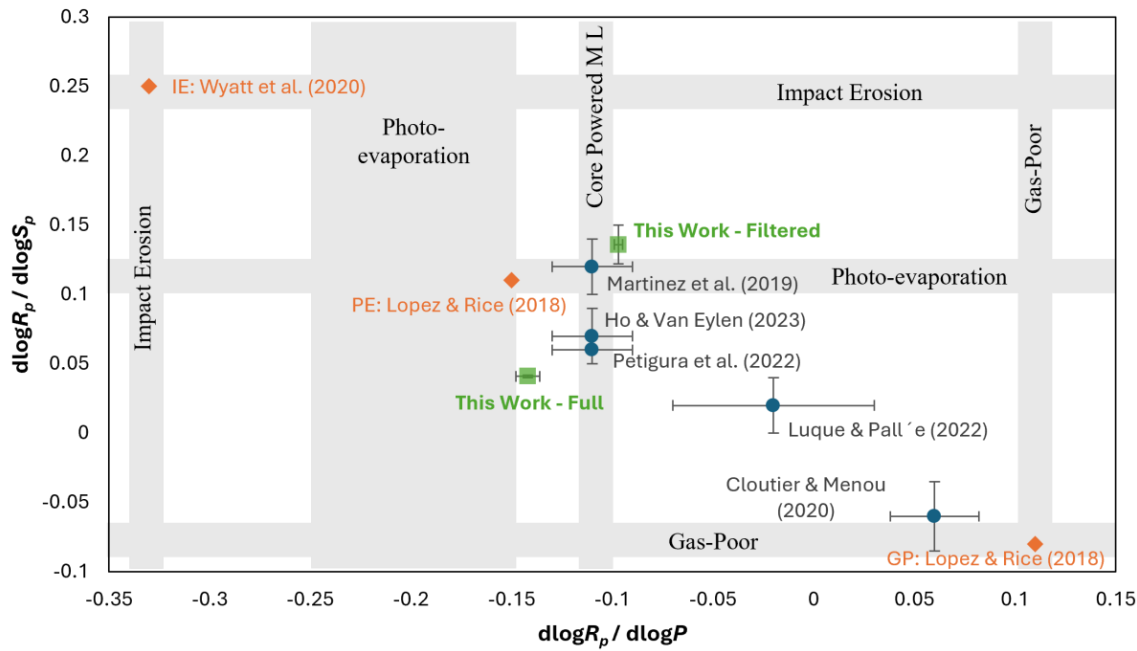

**Figure 9**: Plot of the relative positions of theoretical model predictions from Table 1 (orange diamonds), observational results from Table 2 (blue circles), and this work (green squares) for the slope of the radius valley when viewed as planet radius vs. orbital period (horizontal axis) and planet radius vs. insolation flux (vertical axis) on a log-log scale. Uncertainties are also shown for all observational results including this work. Planet formation processes and their ranges are the shaded regions as predicted by models for each axis.

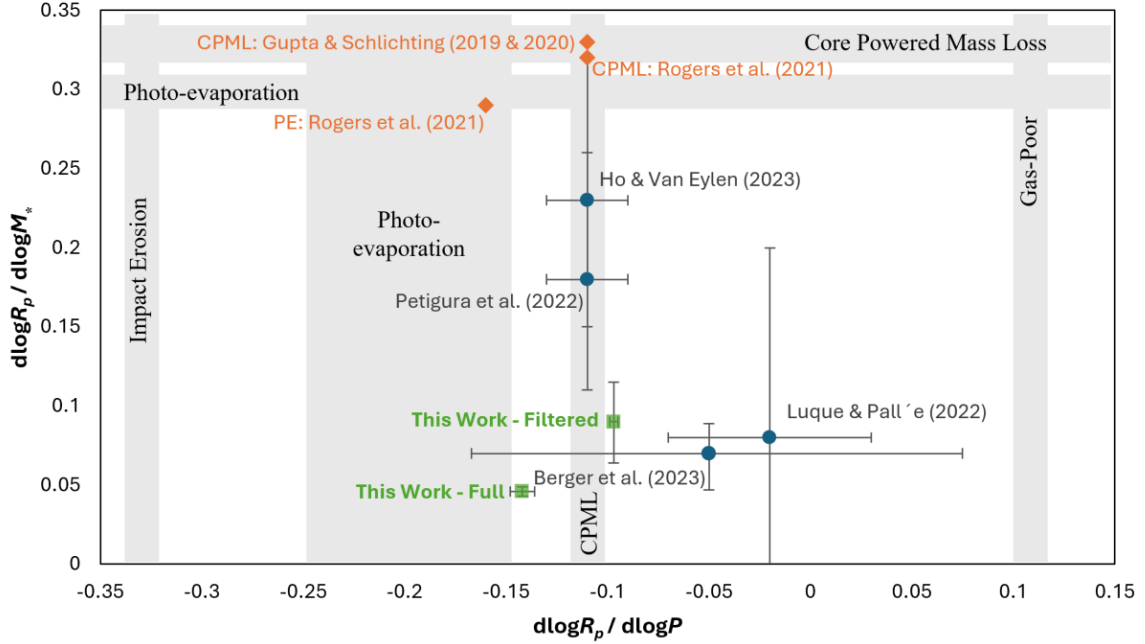

**Figure 10:** Plot of the relative positions of theoretical model predictions from Table 1 (orange diamonds), observational results from Table 2 (blue circles), and this work (green squares) for the slope of the radius valley when viewed as planet radius vs. orbital period (horizontal axis) and planet radius vs. stellar mass (vertical axis) on a log-log scale. Uncertainties are also shown for all observational results including this work. Planet formation processes and their ranges are the shaded regions as predicted by models for each axis.

Taking all this into account, we conclude our results are very consistent with TDML and possibly more consistent with PE than CPML but still inconclusive. This, coupled with some of the bias introduced in trying to maintain a large high-precision sample, makes drawing firm conclusions for PE vs. CPML even more challenging. To facilitate comparing our results from a sample with $R_p$ Median Uncertainty of 3.4%, for a broad group of FGKM host stellar types, with model predictions and previous observations, we have summarized them in Table 5.

**Table 5:** Radius valley slopes from this Work

| Planet Sample Size | Analysis Estimates (slope) | | | Conclusions |
|---|---|---|---|---|
| | $d\log R_p/d\log P$ (R⊕/days) | $d\log R_p/d\log S_p$ (R⊕/S⊕) | $d\log R_p/d\log M_*$ (R⊕/M☉) | |
| 1923 | $-0.142^{+0.006}_{-0.006}$ | $0.041^{+0.001}_{-0.001}$ | $0.046^{+0.003}_{-0.003}$ | Consistent with TDML |
| 1292 | $-0.097^{+0.002}_{-0.002}$ | $0.136^{+0.014}_{-0.014}$ | $0.090^{+0.026}_{-0.025}$ | Consistent with TDML |

# 4. Summary & Conclusions

In this work we assess the predictions of several planet formation models and build on previous research to provide further evidence of planet formation drivers. By enhancing standard SED fitting techniques using Bayesian statistical methods we were able to derive more accurate and precise host star parameters leading to an increase in the accuracy and precision of the associated planet parameters. We show that our enhanced SED fitting methods produced stellar and planetary parameter precisions better than those available for a larger number of exoplanets than in previous studies. This allowed us to produce a sample of 1923 planets with better precision than all previous radius valley studies with over 1000 planets. However, to maintain this large sample in estimating precise planetary parameters we had to use a combination of methods based on the data available from the NASA Archive. This resulted in some parameter values not being completely homogeneous which should be considered in developing conclusions.

By analyzing the radius valley with our larger, more precise planetary radius data, our most conclusive result was consistent with photoevaporation (PE) with a planet radius vs. orbital period slope of $R_p = P^{-0.142^{+0.006}_{-0.006}} e^{0.896^{+0.012}_{-0.010}}$ for the full sample when compared to the various models. For planet radius vs. insolation flux, the filtered sample with a slope of $R_p = S_p^{0.136^{+0.014}_{-0.014}} e^{-0.085^{+0.031}_{-0.030}}$ was also closest to the PE model prediction, but they were slightly different samples from the $R_p$ vs. Period slope results. In comparing our results with past observational studies, our results align better with studies concluding core-powered mass loss such as Ho & Van Eylen (2023) and Berger et al. (2023). Also, the slope as a function of stellar mass for both samples appeared the more consistent with thermally-driven processes when compared to models and previous studies. Additional perspectives of the relative positions of all measurements to all available theoretical and observational data further confirmed some thermally-driven mass loss process as opposed to other models or theories.

## 4.1. Recommended Improvements & Follow Up Research

Even though we were able to derive some conclusions regarding the probability of the planet formation process of the radius valley for this planet sample, we have several recommendations for follow up research and future studies as outlined below.

- Larger Sample Size: Although SEDmc was able to provide a relatively large and precise sample of planets, carefully expanding the pool of exoplanets to include mature candidates as well as confirmed planets would increase

- the final sample size to several thousand. This larger sample size would enable more flexibility in the analysis, reduce the need to use multiple models and methods for parameter estimation and thus more accuracy in the results.
- Parameter Segmentation: Once we have a larger sample size of host stars and planets with precise parameters, we could then perform much better segmentations around stellar mass and other parameters, such as period, insolation flux or age. This will not only allow a closer look at specific types of stars and planets but is also needed to fully utilize some of the new 3D slope measurement techniques outlined in the later studies.
- Improved Slope Measurements: Improvements would include ensuring all assumptions and nuisance parameters are addressed using statistical methods that would eliminate biases and provide additional uncertainties as needed. This along with using the 3D slope measurement techniques enabled by a larger sample size could greatly improve all slope measurements by reducing the sensitivity to sample size and precision when distinguishing between CPML and PE.
- Formation Process Segmentation: Identifying the dominant planet formation process is challenging because there are probably multiple processes involved. In addition, the primary planet formation and evolutionary processes for a particular set of planets or host stars could be different depending on a variety of parameters including period, eccentricity, other planets in the system, as well as the mass, size, age, metallicity, etc. of the host star. More recent studies investigated some of this segmentation with several focused on analyzing the Radius Valley only for M-dwarf stars (Bonfanti et al. 2024; Gaidos et al. 2024; Ho et al. 2024; Venturini et al. 2024). Instead of trying to fit a large group of planets into one dominant process, future studies could look at the specific parameters that drive each process and then try to group planets into categories according to those parameters.

Overall, we feel there are still opportunities to better understand the primary formation and evolutionary process creating the Radius Valley feature between super-Earth and sub-Neptune size planets that are close to their host star. These additional chances for scientific insights will arise through further discoveries, confirmation and precise characterization of exoplanets in our galaxy, along with improved analysis methods verified through continued research in this area.

## Acknowledgements

This research has made use of the VizieR catalog access tool, CDS, Strasbourg, France (DOI: 10.26093/cds/vizier). The original description of the VizieR service was published in 2000, A&AS, 143, 23. This research has made use of the SIMBAD database, CDS, Strasbourg Astronomical Observatory, France published in 2000, A&AS, 143, 9. This work has made use of data from the European Space Agency (ESA) mission Gaia (https://www.cosmos.esa.int/gaia), processed by the Gaia Data Processing and Analysis Consortium (DPAC, https://www.cosmos.esa.int/web/gaia/dpac/consortium). Funding for the DPAC has been provided by national institutions, in particular the institutions participating in the Gaia Multilateral Agreement. This research has made use of the NASA Exoplanet Archive (DOI: 10.26133/NEA13), which is operated by the California Institute of Technology, under contract with the National Aeronautics and Space Administration under the Exoplanet Exploration Program.

*Software:* emcee (Foreman-Mackey et al. 2012), corner (Foreman-Mackey 2016), scipy (Virtanen et al. 2020), astropy (The Astropy Collaboration et al. 2013, 2018, 2022), astroquery (Ginsburg et al. 2019), numpy (Harris et al. 2020)), pandas (The Pandas Development Team 2020), matplotlib (Hunter 2007), scikit-learn (Pedregosa et al. 2011), vizier_sed (Fouesneau 2017), fpdf (Plathey 2004), seaborn (Waskom 2021).

## ORCID IDs

David Jordan  https://orcid.org/0009-0001-1386-2123

Inseok Song  https://orcid.org/0000-0002-5815-7372